\documentclass[pra,twocolumn,showpacs,letterpaper]{revtex4}%
\usepackage{times}
\usepackage{latexsym}
\usepackage{graphicx}
\usepackage{amsmath,amssymb}
\usepackage{verbatim,times,bbm}
\usepackage{amsmath}
\usepackage{amsfonts}
\usepackage{amssymb}
\usepackage{color}
\newtheorem{theorem}{Theorem}
\newtheorem{lemma}{Lemma}
\begin{document}
\title{Quantum Reading Capacity}
\author{Stefano Pirandola}
\thanks{These authors are equal first authors.}
\affiliation{Department of Computer Science, University of York, York YO10 5GH, UK}
\author{Cosmo Lupo}
\thanks{These authors are equal first authors.}
\affiliation{School of Science and Technology, University of Camerino, I-62032 Camerino, Italy}
\author{Vittorio Giovannetti}
\affiliation{NEST, Scuola Normale Superiore and Istituto Nanoscienze-CNR, I-56126 Pisa, Italy}
\author{Stefano Mancini}
\affiliation{School of Science and Technology, University of Camerino, I-62032 Camerino, Italy}
\affiliation{INFN-Sezione di Perugia, I-06123 Perugia, Italy}
\author{Samuel L. Braunstein}
\affiliation{Department of Computer Science, University of York, York YO10 5GH,~UK}

\begin{abstract}
The readout of a classical memory can be modelled as a problem of
quantum channel discrimination, where a decoder retrieves
information by distinguishing the different quantum channels
encoded in each cell of the memory [S. Pirandola, Phys. Rev. Lett.
\textbf{106}, 090504 (2011)]. In the case of optical memories,
such as CDs and DVDs, this discrimination involves lossy bosonic
channels and can be remarkably boosted by the use of nonclassical
light (quantum reading). Here we generalize these concepts by
extending the model of memory from single-cell to multi-cell
encoding. In general, information is stored in a block of cells by
using a channel-codeword, i.e., a sequence of channels chosen
according to a classical code. Correspondingly, the readout of
data is realized by a process of \textquotedblleft
parallel\textquotedblright\ channel discrimination, where the
entire block of cells is probed simultaneously and decoded via an
optimal collective measurement. In the limit of an infinite block
we define the quantum reading capacity of the memory, quantifying
the maximum number of readable bits per cell. This notion of
capacity is nontrivial when we suitably constrain the physical
resources of the decoder. For optical memories (encoding bosonic
channels), such a constraint is energetic and corresponds to
fixing the mean total number of photons per cell. In this case, we
are able to prove a separation between the quantum reading
capacity and the maximum information rate achievable by classical
transmitters, i.e., arbitrary classical mixtures of coherent
states. In fact, we can easily construct nonclassical transmitters
that are able to outperform any classical transmitter, thus
showing that the advantages of quantum reading persist in the
optimal multi-cell scenario.

\end{abstract}

\maketitle

\section{Introduction\label{intro}}

One of the central problems in the field of quantum information is the
statistical discrimination of quantum states~\cite{Helstrom, FuchsTH,
STATE,Minko}. This is a fundamental issue in many protocols, including those
of quantum communication~\cite{Caves,HSW,Holevo,Qcomm} and quantum
cryptography~\cite{Gisin, BB84,Ekert, QKD,Scarani}. A similar problem is the
statistical discrimination of quantum channels, also called \textquotedblleft
quantum channel discrimination\textquotedblright\ (QCD)~\cite{QCD}. In its
basic formulation, QCD\ involves a discrete ensemble of quantum channels which
are associated with some a priori probabilities. A channel is randomly
extracted from the ensemble and given to a party who tries to identify it by
using input states and output measurements. The optimal performance is
quantified by a minimum error probability which is generally non-zero in the
presence of constraints (e.g., for fixed number of queries or restricted space
of the input states). In general, this is a double-optimization problem whose
optimal choices are unknown, a feature which makes its exploration
non-trivial. Moreover QCD\ may also involve continuous ensembles. A special
case is the \textquotedblleft quantum channel estimation\textquotedblright%
\ where the ensemble is indexed by a continuous parameter with flat
distribution. Here the goal is to estimate the unknown parameter with minimal
uncertainty~\cite{CHest,Vitto}.

Besides its difficult theoretical resolution, QCD is also interesting for its
potential practical implementations. For instance, it is at the basis of the
decoding procedure of the two-way quantum cryptography~\cite{QKD2way}\ where
the secret information is encoded in a Gaussian ensemble of phase-space
displacements. Furthermore QCD\ appears also in the quantum illumination of
targets~\cite{Qillumination,Q1}, where the sensing of a remote
low-reflectivity object in a bright thermal environment corresponds to the
binary discrimination between a very noisy/lossy channel (presence of target)
and a completely depolarizing channel (absence of target).

More recently, QCD has been connected with another fundamental
task:\ the readout of classical digital memories~\cite{Pirs}.
Thanks to this connection, Ref.~\cite{Pirs} has laid the basic
ideas of treating digital memories, such as optical disks, in the
field of quantum information theory (see also the following
studies of Refs.~\cite{NairLAST,DARIA3,NairLAST2}). The storage of
data, i.e., the writing of the memory, corresponds to a process of
channel encoding, where information is recorded into a cell by
storing a quantum channel picked from some pre-established
ensemble. Then the process of readout corresponds to the process
of channel decoding, which is equivalent to discriminate between
the various channels of the ensemble. This is done by probing the
cell using an input state, also called \textquotedblleft
transmitter\textquotedblright, and measuring the output by a
suitable detector or \textquotedblleft receiver\textquotedblright.
Ref.~\cite{Pirs} developed this model directly in the bosonic
setting, in order to apply the results to optical memories, such
as CDs and DVDs. The central investigation regarded the comparison
between classical and nonclassical transmitters, where
\textquotedblleft classical transmitters\textquotedblright\
correspond to probabilistic mixtures of coherent states and
encompass all the sources of light which are used in today's data
storage technology. By contrast, \textquotedblleft nonclassical
transmitters\textquotedblright\ are only produced in quantum
optics' labs and they are typically based on entangled, squeezed
or Fock states~\cite{Walls,Knight}. As shown by Ref.~\cite{Pirs},
we can contruct nonclassical transmitters that are able to
outperform any classical transmitter. In particular, this happens
in the regime of low energy, where a few photons are irradiated
over each cell of the memory. This regime is particularly
important for the non trivial implications it can have in terms of
increasing data-transfer rates and storage capacities. Following
the terminology of Ref.~\cite{Pirs}, we call \textquotedblleft
quantum reading\textquotedblright\ the use of nonclassical
transmitters to read data from classical digital memories.

The main results on the quantum reading of memories regarded the single-cell
scenario, where each memory cell is written and read independently from the
others. However, a supplementary analysis of Ref.~\cite{Pirs} also showed that
the advantages of quantum reading persist when we extend the encoding of
information from a single- to a multi-cell model. Assuming a block-encoding of
data, one can use error correcting codes which make the readout flawless. In
this scenario it is possible to show that the error correction overhead can be
made negligible at low energies only when we adopt nonclassical
transmitters~\cite{PirsSUPP}. Motivated by this analysis, the present work
provides a full general treatment of the quantum reading of memories in the
multi-cell scenario. This is done by formalizing the most general kind of
classical digital memory. In this model, information is stored in a block of
cells by using a channel-codeword, i.e., a sequence of channels chosen
according to some classical code. Then, the readout of data is realized by a
process of parallel channel discrimination. This means that the entire block
of cells is probed in parallel and then decoded by an optimal collective
measurement. Such a description encompasses all the possible encoding and
decoding strategies. Since the storage capacity of classical memories is
usually very large, an average memory is made by a large number of these
encoding blocks. The optimal scenario corresponds to the case where the whole
memory is represented by a single, very large, encoding block which is read in
a parallel fashion. In this limit (infinite block) we can provide a simple
characterization of the memory and resort to the Holevo bound to quantify the
amount of readable information. This enables us to define the quantum reading
capacity of the classical memory, which corresponds to the maximum
\textit{readable} information per cell. If we do not impose constraints, this
capacity equals exactly the amount of information stored in each cell of the
memory. However, this is no longer the case when we introduce physical
constraints on the resources accessible to the reading device. In the case of
optical memories, which involve the discrimination of bosonic channels, the
energy constraint is the most fundamental~\cite{Caves}. Thus the quantum
reading capacity is properly formulated for fixed input energy. This means
that we fix the mean total number of photons irradiated over each cell of the
memory. The computation of this capacity would be very important at the low
energy regime, which is the most interesting for its potential implications.
Despite its calculation is extremely difficult, we are able to provide lower
bounds for the most basic optical memories, i.e., the ones based on the binary
encoding of lossy channels. For these memories we are able to derive a simple
lower bound which quantifies the maximum information readable by classical
transmitters. We call this bound the \textquotedblleft classical reading
capacity\textquotedblright\ of the memory and represents an extension to the
multi-cell scenario of the \textquotedblleft classical discrimination
bound\textquotedblright\ introduced in Ref.~\cite{Pirs}. Remarkably, the
optimal classical transmitter which irradiates $n$ mean photons per cell can
be realized by using a single coherent state with the same mean number of
photons. Thanks to this result, we can easily investigate if a particular
nonclassical transmitter is able to outperform any classical transmitter. This
is indeed what we find in the regime of few photons. Thus, in the low energy
regime, we can prove the separation between the quantum reading capacity and
the classical reading capacity, which is equivalent to state that the
advantages of quantum reading persist in the optimal multicell scenario.

The paper is organized as follows. In Secs.~\ref{singleSECTION}
and~\ref{SimpleMemorySEC} we review some of the key-points of Ref.~\cite{Pirs}
and its supplementary materials, which are preliminary for the new results of
Secs.~\ref{MulticellSEC}-\ref{NonclassicalSEC}. In particular, in
Sec.~\ref{singleSECTION}, we review the basic notions regarding the memory
model with single-cell encoding. Then, in the following
Sec.~\ref{SimpleMemorySEC}, we discuss the simplest example of optical memory
and its quantum reading. Once we have reviewed these notions, we introduce the
model with multi-cell encoding in Sec.~\ref{MulticellSEC}. In
Sec.~\ref{InfiniteSEC} we take the limit for infinite block size and we define
the quantum reading capacity of the memory, both unconstrained and
constrained. In particular, we specialize the constrained capacity to the case
of optical memories (bosonic channels). In Sec.~\ref{BoundSEC} we compute the
lower bound relative to classical transmitters, i.e., the classical reading
capacity. In the following Sec.~\ref{NonclassicalSEC} we prove that this bound
is separated, by showing simple examples of nonclassical transmitters which
outperform classical ones in the regime of few photons. Finally,
Sec.~\ref{SEC_Conclusion} is for conclusions.

\section{Basic model of memory: single-cell encoding \label{singleSECTION}}

\begin{figure}[ptbh]
\vspace{-0.7cm}
\par
\begin{center}
\includegraphics[width=0.5\textwidth] {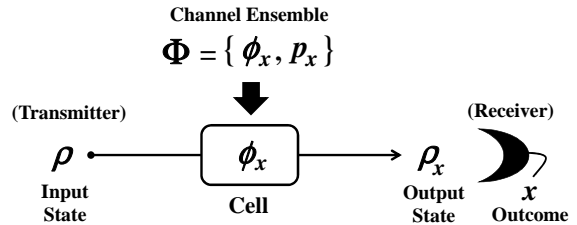}
\end{center}
\par
\vspace{-1.8cm} \caption{\textbf{Basic process of storage and
readout.} A memory cell can be characterized by an ensemble of
quantum channels $\Phi=\left\{  \phi_{x},p_{x}\right\}  $. Alice
picks a quantum channel $\phi_{x}$\ (with probability $p_{x}$) and
stores it in a target cell. In order to read the information, Bob
exploits a transmitter and a receiver. In the simplest scenario,\
this corresponds to inputting a suitable quantum state $\rho$ and
measuring the output $\rho_{x}=\phi_{x}(\rho)$ by a\ suitable
detector. The detector gives the correct answer $x$\ up to some
error probability $P_{err}$. \textbf{Multi-copy probing.}~Since
the cell encodes the quantum channel in a stable way, we can probe
the cell many times. This means that, more generally, Bob can
input a multipartite state $\rho(s)\in
\mathcal{D}(\mathcal{H}^{\otimes s})$ which describes $s$\ quantum
systems. As a consequence, the output will be
$\rho_{x}(s)=\phi_{x}^{\otimes s}[\rho(s)]$, whose global
detection gives $x$ up to an error probability (which is
non-increasing in $s$). \textbf{Optical memory. }The encoded
channel $\phi _{x}$ is a bosonic channel (in particular,
single-mode). In this case, Bob uses an input state $\rho(s,n)$
describing $s$ bosonic modes and irradiating $n$ mean photons over
the cell.}\label{Fig1}
\end{figure}

In the more abstract sense, a classical digital memory can be modelled as a
one-dimensional array of cells (the generalization to two or more dimensions
is just a matter of technicalities). The writing of information by some device
or encoder, that we just call \textquotedblleft Alice\textquotedblright\ for
simplicity, can be modelled as a process of channel encoding~\cite{Pirs}. This
means that Alice has a classical random variable $X=\{x,p_{x}\}$ with
$k$\ values $x=0,\cdots,k-1$ distributed according to a probability
distribution $p_{x}$. Each value $x$ is then associated with a quantum
channel\ $\phi_{x}$\ via one-to-one correspondence%
\begin{equation}
x\leftrightarrow\phi_{x}~,
\end{equation}
thus defining an ensemble of quantum channels
\begin{equation}
\Phi=\left\{  \phi_{x},p_{x}\right\}  ~.
\end{equation}
Mathematically speaking, each channel of the ensemble is a completely positive
trace-preserving (CPT) map acting on the state space $\mathcal{D}%
(\mathcal{H})$ of some chosen quantum system (Hilbert space $\mathcal{H}$).
Furthermore, the various channels are different from each other. This means
that, for any pair $\phi_{x}$ and $\phi_{x^{\prime}}$, there is at least one
state $\rho\in\mathcal{D}(\mathcal{H})$ such that
\begin{equation}
F[\phi_{x}(\rho),\phi_{x^{\prime}}(\rho)]<1~,
\end{equation}
where
$F(\rho,\sigma)=[\mathrm{Tr}(\sqrt{\rho}\sigma\sqrt{\rho})^{1/2}]^{2}$
is the quantum fidelity~\cite{Fidelity}. Thus, in order to write
information, Alice randomly picks a quantum channel $\phi_{x}$
from the ensemble and stores it in a target cell. This operation
is repeated identically and independently for all the cells of the
memory, so that we can characterize both the cell and the memory
by specifying $\Phi$ (see Fig.~\ref{Fig1}).

The readout of information corresponds to the inverse process, which is
channel decoding or discrimination. The written memory is passed to a decoder,
that we call \textquotedblleft Bob\textquotedblright, who queries the cells of
the memory one by one. To retrieve information from a target cell, Bob
exploits a transmitter and a receiver. In the simplest case this means that
Bob inputs a suitable quantum state $\rho$ and measures the corresponding
output state $\rho_{x}=\phi_{x}(\rho)$ recording the specific quantum channel
stored in that cell (see Fig.~\ref{Fig1}). Note that, given some input state
$\rho$, the ensemble of the possible output states $\{\phi_{x}(\rho),p_{x}\}$
is generally made by non-orthogonal states which, therefore, cannot be
perfectly distinguished by a quantum measurement. In other words, the
discrimination cannot be perfect and the quantum detection will output the
correct value $x$\ up to an error probability $P_{err}$. It is clear that the
main goal for Bob is to optimize input state and output measurement in order
to retrieve the maximal information from the cell.

\subsection{Multi-copy probing and optical memories}

In a classical digital memory information is stored (quasi)permanently. This
means that the association between a single cell and the channel-encoding
$\Phi$ must be stable. As a result, Bob can probe the cell many times by using
an input state living in a bigger state space. Given some quantum channel%
\begin{equation}
\phi_{x}:\mathcal{D}(\mathcal{H})\rightarrow\mathcal{D}(\mathcal{H})~,
\end{equation}
Bob can input a multipartite state $\rho(s)\in\mathcal{D}(\mathcal{H}^{\otimes
s})$ with integer $s\geq1$, i.e., describing $s$\ quantum systems. As a
result, the output state will be
\begin{equation}
\rho_{x}(s)=\phi_{x}^{\otimes s}[\rho(s)]~.
\end{equation}
This state is detected by a quantum measurement applied to the whole set of
$s$ quantum systems (see Fig.~\ref{Fig1}). Physically, if we consider the
process in the time domain, $\rho(s)$ describes the global state of
$s$\ systems which are \textit{sequentially} transmitted through the cell. In
other words, the number $s$ can also be regarded as a dimensionless readout
time~\cite{Pirs}. Intuitively, it is expected that the optimal $P_{err}$ is a
decaying function of $s$, so that it is always possible to retrieve all the
information in the limit for $s\rightarrow\infty$. This suggests that the
readout problem is nontrivial only if we impose constraints on the physical
resources that are used to probe the memory. In the case of discrete variables
(i.e., finite-dimensional Hilbert space) the constraint can be stated in terms
of fixed or maximum readout time $s$.

More fundamental constraints come into play when we consider an optical
memory, which can be defined as classical memory encoding an ensemble $\Phi
$\ of bosonic channels. In particular, these channels can be assumed to be
single-mode. Since the underlying Hilbert space is infinite in the bosonic
setting, one has unbounded operators such as the energy. Clearly, if we allow
the energy to go to infinite, the discrimination of (different) bosonic
channels is always perfect. As a result, the readout of optical memories has
to be modelled as a channel discrimination problem where we fix the input
energy. The minimal energy constraint corresponds to fixing the mean total
number of photons $n$ irradiated \textit{over} each memory cell~\cite{Pirs}.
Thus, for fixed $n$, the aim of Bob is to optimize input (i.e., number of
bosonic systems $s$ and their state $\rho$) and the output measurement. In the
following we explicitly formalize this constrained problem.

Let us consider an optical memory with cell $\Phi=\left\{  \phi_{x}%
,p_{x}\right\}  $ where each element $\phi_{x}$ is a single-mode bosonic
channel. Then, we denote by $\rho(s,n)$ a multimode bosonic state $\rho
\in\mathcal{D}(\mathcal{H}^{\otimes s})$ with mean total energy \textrm{Tr}%
$(\rho\hat{n})=n$, where $\hat{n}$ is the total number operator over
$\mathcal{H}^{\otimes s}$. In other words, this state describes $s$ bosonic
systems which irradiate a total of $n$ mean photons over the target cell (see
also Fig.~\ref{Fig1}). We refer to the pair $(s,n)$ as to the signal profile.
In the bosonic setting the parameter $s$ can be interpreted not only as the
number of \textit{temporal} modes (therefore, readout time) but equivalently
as the number of \textit{frequency} modes, thus quantifying the
\textquotedblleft bandwidth\textquotedblright\ of the signal~\cite{Pirs}. Now,
for a given input $\rho=\rho(s,n)$ to the cell $\Phi$, we have the output
state%
\begin{equation}
\rho_{x}(s,n)=\phi_{x}^{\otimes s}[\rho(s,n)]~. \label{rhoX}%
\end{equation}
This output is subject to a quantum measurement over the $s$ modes which is
generally described by a positive operator valued measure (POVM)
$\mathcal{M}=\{\Pi_{x}\}$ having $k$ detection operators $\Pi_{x}\geq0$ which
sums up to the identity $\sum_{x}\Pi_{x}=I$. This measurement gives the
correct answer $x$ up to an error probability%
\begin{equation}
P=1-\sum_{x=0}^{k-1}p_{x}\mathrm{Tr}[\Pi_{x}\rho_{x}(s,n)]:=P[\Phi
|\rho(s,n),\mathcal{M}]~. \label{Pfirst}%
\end{equation}
Here we denote by $P[\Phi|\rho(s,n),\mathcal{M}]$ the error probability in the
readout of the cell $\Phi$ given an input state $\rho(s,n)$ and an output
measurement $\mathcal{M}$. Now we are interested in minimizing this quantity
over input and output.

As a first step we fix the signal profile $(s,n)$ and consider the
minimization over input states and output measurements. This leads to the quantity%

\begin{equation}
P(\Phi|s,n)=\min_{\rho(s,n),\mathcal{M}}P[\Phi|\rho(s,n),\mathcal{M}]~,
\label{Psecond}%
\end{equation}
which is the minimum error probability achievable for a fixed signal profile
$(s,n)$. Note that there are some cases where the optimal output POVM\ is
known. For instance if the output states $\rho_{x}(s,n)$ are pure and form a
geometrically uniform set~\cite{Ban,Kato}, then the optimal detection is the
square root measurement~\cite{Helstrom}.

As a final step, we keep the energy $n$\ fixed and we minimize over $s$, thus
defining the minimum error probability at fixed energy per cell, i.e.,%
\begin{equation}
P(\Phi|n)=\inf_{s}P(\Phi|s,n)~. \label{OptimalP}%
\end{equation}
Thus, given a memory with cell $\Phi$, the determination of $P(\Phi|n)$
provides the \textquotedblleft optimal\textquotedblright\ readout of the cell
at fixed energy $n$. It is worth stressing that the minimization over the
number of signals $s$ is not trivial due to the constraint that we impose on
the mean total energy (if instead of such restriction one imposes a bound on
the mean energy \textit{per signal}, then the infimum is always achieved in
the asymptotic limit of $s\rightarrow\infty$). Also notice that we have put
the word \textquotedblleft optimal\textquotedblright\ between apostrophes,
since the optimality of Eq.~(\ref{OptimalP}) is still partial, i.e., not
including all the possible readout strategies. In fact, as we discuss in the
following subsection, Bob can also consider the help of ancillary systems
while keeping equal to $n$ the mean total number of photons irradiated over
the cell.

\subsection{Assisted readout of optical memories}

\begin{figure}[ptbh] \vspace{-0.4cm}
\par
\begin{center}
\includegraphics[width=0.5\textwidth] {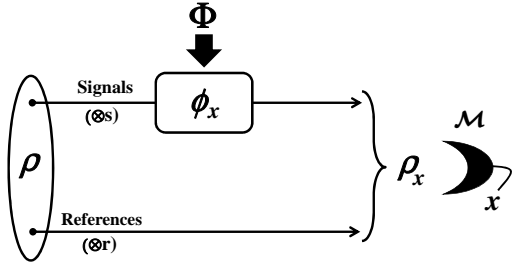}
\end{center}
\par
\vspace{-2.1cm} \caption{\textbf{Assisted readout of an optical
memory.} Alice stores data in the cell by encoding a single-mode
bosonic channel $\phi_{x}$ picked from the ensemble $\Phi$. In
general, Bob queries the cell by using a transmitter $\rho(s,r,n)$
which describes $s$ signal modes, irradiating $n$ mean photons
over the cell, plus $r$ reference modes (bypassing the cell). The
global output state $\rho_{x}(s,r,n)$ is detected by a quantum
measurement $\mathcal{M}$, which provides the correct answer $x$
up to an error probability
$P[\Phi|\rho(s,r,n),\mathcal{M}]$.}\label{Fig2}
\end{figure}

The optimality of Eq.~(\ref{OptimalP}) is true only in the \textquotedblleft
unassisted case\textquotedblright\ where all the input modes are sent through
the target cell. More generally, Bob can exploit an interferometric-like setup
by introducing an ancillary \textquotedblleft reference\textquotedblright%
\ system which bypasses the cell and assists the output measurement as
depicted in Fig.~\ref{Fig2}. In the \textquotedblleft assisted
case\textquotedblright\ we consider an input state $\rho\in\mathcal{D}%
(\mathcal{H}_{S}^{\otimes s}\otimes\mathcal{H}_{R}^{\otimes r})$ which
describes $s$ signal modes (Hilbert space $\mathcal{H}_{S}^{\otimes s}$)\ plus
a reference bosonic system with $r$ modes (Hilbert space $\mathcal{H}%
_{R}^{\otimes r}$)~\cite{Assisted}. As before, the minimal energy constraint
corresponds to fixing the mean total number of photons irradiated over the
target cell, i.e., $n=$\textrm{Tr}$(\rho\hat{n}_{S})$ where $\hat{n}_{S}$ is
the total number operator acting over $\mathcal{H}_{S}^{\otimes s}%
$~\cite{NoteEnergy}. We denote by $\rho=\rho(s,r,n)$ such a state, where we
make explicit the number of signal modes $s$, the number of reference modes
$r$, and the mean total number of photons $n$ irradiated over the cell.
Following the language of Ref.~\cite{Pirs}, we also refer to $\rho(s,r,n)$ as
to a transmitter\ with $s$ signals, $r$ references, and signalling $n$
photons~\cite{Transmitters}.

Now, given a transmitter $\rho(s,r,n)$ at the input of a target cell
$\Phi=\left\{  \phi_{x},p_{x}\right\}  $, we have the output state%
\begin{equation}
\rho_{x}(s,r,n)=(\phi_{x}^{\otimes s}\otimes I^{\otimes r})\rho(s,r,n)~,
\label{rhoDILATED}%
\end{equation}
where the channel $\phi_{x}$ acts on each signal mode, while the identity $I$
acts on each reference mode. This state is then measured by a
POVM\ $\mathcal{M}=\{\Pi_{x}\}$ where $\Pi_{x}$ acts on the whole state space
$\mathcal{D}(\mathcal{H}_{S}^{\otimes s}\otimes\mathcal{H}_{R}^{\otimes r})$.
The error probability $P[\Phi|\rho(s,r,n),\mathcal{M}]$ has the form of
Eq.~(\ref{Pfirst}) where now both state and measurement are dilated to the
reference system. Thus, given a memory with cell $\Phi$, the minimum error
probability at fixed signal energy $n$ is given by%
\begin{equation}
P(\Phi|n)=\inf_{s,r}\left\{  \min_{\rho(s,r,n),\mathcal{M}}P[\Phi
|\rho(s,r,n),\mathcal{M}]\right\}  ~, \label{Passisted}%
\end{equation}
where the minimization includes the reference system too. In
general, we always consider the assisted scheme and the
corresponding error probability of Eq.~(\ref{Passisted}). This
clearly represents a superior strategy for the possibility of
using entanglement between signal and reference systems. Clearly,
the unassisted strategy is achieved back by setting $r=0$ and
$\rho(s,0,n)=\rho(s,n)$. 

\section{The simplest case: optical memory with binary
cells\label{SimpleMemorySEC}}

In general the solution of Eq.~(\ref{Passisted}) is extremely difficult. In
order to investigate the problem, the simplest possible scenario corresponds
to an optical memory whose cell encodes two bosonic channels (binary
cell)~\cite{Pirs}. The situation is particularly advantageous when the
channels are pure-loss and they are chosen with the same probability. This
means to consider the binary channel ensemble
\begin{equation}
\bar{\Phi}=\{\phi_{u},p_{u}\}_{u=0,1}=\{\phi_{0},p_{0},\phi_{1},p_{1}\}~,
\end{equation}
where $p_{0}=p_{1}=1/2$, and $\phi_{u}$ represents a pure-loss channel with
transmission $0\leq\kappa_{u}\leq1$. In the Heisenberg picture, the action of
$\phi_{u}$ on each signal mode is given by the map
\begin{equation}
\hat{a}_{S}\rightarrow\sqrt{\kappa_{u}}\hat{a}_{S}-\sqrt{1-\kappa_{u}}\hat
{a}_{E}~,
\end{equation}
where $\hat{a}_{S}$ is the annihilation operator of the signal mode and
$\hat{a}_{E}$ is the one of an environmental mode which is prepared in the
vacuum state. For simplicity we can also denote the ensemble by using the
short notation%
\begin{equation}
\bar{\Phi}=\{\kappa_{0},\kappa_{1}\}~.
\end{equation}
When the optical memory is read in reflection (which is usually the case),
then the two parameters $\kappa_{0}$ and $\kappa_{1}$ represent the two
possible reflectivities of the cell (so that unit reflectivity corresponds to
perfect transmission of the signal from transmitter to receiver).

Given a transmitter$\ \rho(s,r,n)$ at the input of the binary cell $\bar{\Phi
}$, we have two equiprobable outputs, $\rho_{0}(s,r,n)$ and $\rho_{1}(s,r,n)$.
In this case the optimal measurement corresponds to the projection onto the
positive part of the Helstrom matrix $\rho_{0}(s,r,n)-\rho_{1}(s,r,n)$%
~\cite{Helstrom}. As a result, the error probability for reading the binary
cell $\bar{\Phi}$\ using the transmitter $\rho(s,r,n)$ is given by%
\begin{equation}
P[\bar{\Phi}|\rho(s,r,n)]=\frac{1}{2}\left\{  1-\frac{1}{2}D[\rho
_{0}(s,r,n),\rho_{1}(s,r,n)]\right\}  ~,
\end{equation}
where $D$\ is the trace distance~\cite{Helstrom}. This expression has to be
optimized on the input only, so that we can write%
\begin{equation}
P(\bar{\Phi}|n)=\inf_{s,r}\left\{  \min_{\rho(s,r,n)}P[\bar{\Phi}%
|\rho(s,r,n)]\right\}  ~, \label{Pbinary}%
\end{equation}
which is the minimum error probability at fixed signal energy. This quantity
clearly provides the maximum information per cell at fixed signal energy,
which is given by%
\begin{equation}
I(\bar{\Phi}|n)=1-H[P(\bar{\Phi}|n)]~,
\end{equation}
where
\begin{equation}
H(x)=-x\log_{2}{x}-(1-x)\log_{2}{(1-x)}%
\end{equation}
is the binary formula of the Shannon entropy.

Even in this simple binary case the solution of Eq.~(\ref{Pbinary}) is very
difficult. However we can provide remarkable lower bounds if we restrict the
minimization to some suitable class of transmitters. An important class is the
one of the classical transmitters, since they encompass all the optical
resources used for the readout of optical memories in today's storage
technology. Furthermore, this class can be easily characterized.\ Given a
transmitter $\rho(s,r,n)$, we can write its Glauber-Sudarshan
representation~\cite{Suda}%
\begin{equation}
\rho(s,r,n)=\int d^{2s}\alpha~d^{2r}\beta~P(\alpha,\beta)~\sigma
(\alpha)\otimes\gamma(\beta)~,
\end{equation}
where $\alpha=(\alpha_{1},\cdots,\alpha_{s})^{T}$ and $\beta=(\beta_{1}%
,\cdots,\beta_{r})^{T}$ are vectors of complex amplitudes,%
\begin{equation}
\sigma(\alpha)=\bigotimes_{i=1}^{s}|\alpha_{i}\rangle_{S}\langle\alpha
_{i}|~,~\gamma(\beta)=\bigotimes_{i=1}^{r}|\beta_{i}\rangle_{R}\langle
\beta_{i}|~,
\end{equation}
are multimode coherent states, and the $P$-function $P(\alpha,\beta)$ is a
quasi-distribution, i.e., normalized to one but generally non
positive~\cite{Suda}. In terms of the $P$-function, the signal energy
constraint reads
\begin{equation}
\int d^{2s}\alpha d^{2r}\beta~P(\alpha,\beta)\sum_{i=1}^{s}|\alpha_{i}%
|^{2}=n~. \label{PR-constraint}%
\end{equation}
Now we say that $\rho(s,r,n)$ is classical (nonclassical) if the $P$-function
is positive (non positive). Thus if the transmitter is classical, denoted by
$\rho_{c}(s,r,n)$, then it can be represented as a probabilistic mixture of
coherent states. The simplest examples of classical transmitters are the
coherent state transmitters, that we denote by $\rho_{coh}(s,r,n)$. These are
defined by singular $P$-functions
\begin{equation}
P(\alpha,\beta)=\delta^{2s}(\alpha-\bar{\alpha})\delta^{2r}(\beta-\bar{\beta
})~,
\end{equation}
so that they have the simple form%
\[
\rho_{coh}(s,r,n)=\sigma(\bar{\alpha})\otimes\gamma(\bar{\beta})~.
\]
Examples of nonclassical transmitters are constructed using squeezed states,
entangled states and number states~\cite{Walls,Knight}.

As shown in Ref.~\cite{Pirs}, by restricting the optimization to classical
transmitters, we can compute the upper bound%
\begin{equation}
P(\bar{\Phi}|n)\leq P_{c}(\bar{\Phi}|n):=\inf_{s,r}\left\{  \min_{\rho
_{c}(s,r,n)}P[\bar{\Phi}|\rho_{c}(s,r,n)]\right\}  ,
\end{equation}
which is given by
\begin{equation}
P_{c}(\bar{\Phi}|n)=\frac{1-\sqrt{1-\exp[-n(\sqrt{\kappa_{0}}-\sqrt{\kappa
_{1}})^{2}]}}{2}~. \label{classBOUND}%
\end{equation}
This bound can be reached by a coherent state transmitter $\rho_{coh}%
(1,0,n)=|\sqrt{n}\rangle_{S}\langle\sqrt{n}|$, i.e., a single-mode coherent
state with mean number of photons equal to $n$. The error probability
$P_{c}(\bar{\Phi}|n)$ of Eq.~(\ref{classBOUND}) or, equivalently, the mutual
information
\begin{equation}
I_{c}(\bar{\Phi}|n)=1-H[P_{c}(\bar{\Phi}|n)]~, \label{classBOUNDinfo}%
\end{equation}
is known as \textquotedblleft classical discrimination bound\textquotedblright.

Alternative (and better) bounds can be derived by resorting to nonclassical
transmitters~\cite{Pirs}. As a prototype of nonclassical transmitter we
consider the EPR\ transmitter~\cite{Qillumination}, which is composed by $s$
pairs of signals and references, entangled via two-mode squeezing. This
transmitter has the form
\begin{equation}
\rho_{epr}(s,s,n)=\left\vert \xi\right\rangle \left\langle \xi\right\vert
^{\otimes s}%
\end{equation}
where $\left\vert \xi\right\rangle \left\langle \xi\right\vert $ is a two mode
squeezed vacuum (TMSV) state, entangling one signal mode $S$ with one
reference mode $R$. In the number-ket representation, we have~\cite{Walls}
\begin{equation}
\left\vert \xi\right\rangle =(\cosh\xi)^{-1}\sum_{m=0}^{\infty}(\tanh\xi
)^{m}\left\vert m\right\rangle _{S}\left\vert m\right\rangle _{R}~,
\end{equation}
where the squeezing parameter $\xi$ quantifies the signal-reference
entanglement and gives the energy of the signal by $\sinh^{2}\xi$. Since this
transmitter involves $s$ copies of this state, we have to impose
\begin{equation}
\xi=\mathrm{arc}\sinh\sqrt{\frac{n}{s}}~,
\end{equation}
in order to have an average of $n$ total photons irradiated over the cell.
Given an EPR transmitter $\rho_{epr}(s,s,n)$ at the input of the binary cell
$\bar{\Phi}$, we have an error probability $P[\bar{\Phi}|\rho_{epr}(s,s,n)]$.
By optimizing over the number of copies $s$, we define the upper bound%
\begin{equation}
P(\bar{\Phi}|n)\leq P_{epr}(\bar{\Phi}|n):=\inf_{s}P[\bar{\Phi}|\rho
_{epr}(s,s,n)]~.
\end{equation}
This bound represents the maximum information which can be read from the
binary cell $\bar{\Phi}$ by using an EPR\ transmitter which signal $n$ mean
photons. This quantity can be estimated using the quantum Battacharyya bound
and its Gaussian formula~\cite{Minko}. After some algebra, we
get~\cite{Pirs,PirsSUPP}%
\begin{equation}
P_{epr}(\bar{\Phi}|n)\leq B:=\frac{\exp(-\omega n)}{2}~, \label{Batta}%
\end{equation}
where%
\begin{equation}
\omega:=\frac{\kappa_{0}+\kappa_{1}+2}{2}-2\sqrt{\kappa_{0}\kappa_{1}}%
-\sqrt{(1-\kappa_{0})(1-\kappa_{1})}~. \label{y_exp}%
\end{equation}

\subsection{Quantum versus classical reading}

Because of the potential implications in information technology, it is
important to compare the performances of classical and nonclassical
transmitters. The basic question to ask is the following~\cite{Pirs}: for
fixed signal energy $n$ irradiated over a binary cell $\bar{\Phi}$, can we
find some EPR\ transmitter able to outperform any classical transmitter? In
other words, this is equivalent to show that $P_{epr}(\bar{\Phi}|n)<P_{c}%
(\bar{\Phi}|n)$ and a sufficient condition corresponds to prove that
$B<P_{c}(\bar{\Phi}|n)$. Thus, by using Eqs.~(\ref{classBOUND})
and~(\ref{Batta}), we find that for signal energies
\begin{equation}
n>n_{th}:=\frac{2\ln2}{2-\kappa_{0}-\kappa_{1}-2\sqrt{(1-\kappa_{0}%
)(1-\kappa_{1})}}, \label{nTHRESHOLD}%
\end{equation}
it is always possible to beat classical transmitters by using an
EPR\ transmitter~\cite{Pirs}. For high reflectivity $\kappa_{1}\simeq1$ and
$\kappa_{0}<\kappa_{1}$ the threshold energy $n_{th}$ can be very low. In the
case of \textquotedblleft ideal memories\textquotedblright, defined by
$\kappa_{0}<\kappa_{1}=1$, the bound of Eq.~(\ref{Batta}) can be improved. In
fact, we can write%
\begin{equation}
P_{epr}(\bar{\Phi}|n)\leq\theta:=\frac{\exp[-2n(1-\sqrt{\kappa_{0}})]}{2}~,
\label{thetaINI}%
\end{equation}
and the threshold energy becomes $n_{th}=1/2$~\cite{Pirs}. Thus, for optical
memories with high reflectivities and signal energies $n>1/2$, there always
exists a nonclassical transmitter able to beat any classical transmitter. In
the few-photon regime, roughly given by $1/2<n<10^{2}$, the advantages of
quantum reading can be numerically remarkable, up to one bit per cell. The
implications have been thoroughly discussed in Refs.~\cite{Pirs,PirsSUPP}. It
is important to say that these advantages are also preserved if thermal noise
is added to the basic model. This noise can describe the effect of stray
photons hitting the memory from the background and other decoherence processes
occurring in the reading device. Formally, this means to extend the problem
from the discrimination of pure-loss channels to the discrimination of more
general Gaussian channels~\cite{PirsSUPP}.

A supplementary analysis of quantum reading has also shown that its advantages
persist if we consider more advanced designs of memories where information is
written on and read from block of cells (multi-cell/block encoding). Block
encoding allows Alice to introduce error correcting codes which make Bob's
readout flawless up to some metadata overhead. By resorting to the Hamming
bound and the Gilbert-Varshamov bound, Ref.~\cite{PirsSUPP} showed that
EPR\ transmitters enable the low-energy flawless readout of classical memories
up to a negligible error correction overhead, contrarily to what happens by
employing classical transmitters. In the following section, we develop the
idea of block encoding in the most general scenario, i.e., for arbitrary
classical memories. Then, by sending the size of the block to infinite
(Sec.~\ref{InfiniteSEC}), we will be able to introduce the notion of quantum
reading capacity of a classical memory.

\section{General model of memory: multi-cell encoding\label{MulticellSEC}}

The writing of a memory is based on channel encoding which generally may
involve a block of $m$ cells. A first trivial kind of block encoding is just
based on independent and identical extractions. As usual, Alice encodes a
$k$-ary variable $X=\left\{  x,p_{x}\right\}  $ into an ensemble of quantum
channels $\Phi=\left\{  \phi_{x},p_{x}\right\}  $. Then, she performs $m$
independent extractions from $X$, generating an $m$-letter sequence%
\begin{equation}
\mathbf{x}:=(x_{1},\cdots,x_{m})~,
\end{equation}
with probability $p_{\mathbf{x}}=p_{x_{1}}\cdots p_{x_{m}}$. This classical
sequence identifies a corresponding \textquotedblleft
channel-sequence\textquotedblright\
\begin{equation}
\phi_{\mathbf{x}}:=\phi_{x_{1}}\otimes\cdots\otimes\phi_{x_{m}}~,
\label{chSEQ}%
\end{equation}
which is stored in the block of $m$\ cells.

In a more general approach, Alice adopts a classical code. This means that
Alice disposes a set of $m$-letter codewords $\{\mathbf{x}^{0},\cdots
,\mathbf{x}^{i},\cdots,\mathbf{x}^{l-1}\}$ with $l\leq k^{m}$. A given
codeword
\begin{equation}
\mathbf{x}^{i}=(x_{1}^{i},\cdots,x_{m}^{i})~,
\end{equation}
is chosen with some probability $p_{\mathbf{x}^{i}}$ and identifies a
corresponding \textquotedblleft channel-codeword\textquotedblright\
\begin{equation}
\phi_{\mathbf{x}^{i}}=\phi_{x_{1}^{i}}\otimes\cdots\otimes\phi_{^{x_{m}^{i}}%
}~. \label{channelCODE}%
\end{equation}
Thus, in general, Alice encodes information in a block of $m$ cells by storing
a channel-codeword, which is randomly chosen from the ensemble $\{\phi
_{\mathbf{x}^{i}},p_{\mathbf{x}^{i}}\}$ where $i=0,\cdots,l-1$%
.\begin{figure}[ptbh]
\vspace{-0.2cm}
\par
\begin{center}
\includegraphics[width=0.5\textwidth] {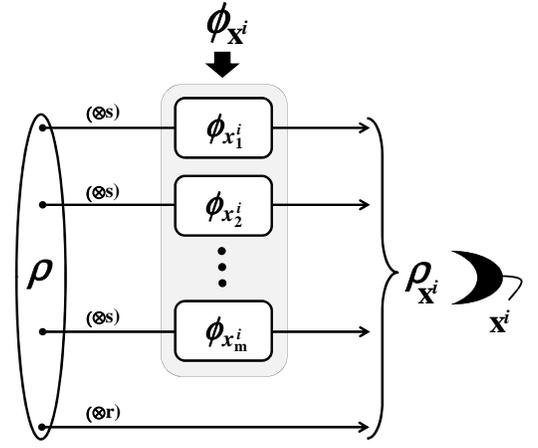}
\end{center}
\par
\vspace{-0.4cm} \caption{\textbf{Memory model with block
encoding.} In order to write data, Alice encodes a
channel-codeword $\phi_{\mathbf{x}^{i}}$ in a block of $m$ cells.
To read the data, Bob uses a suitable transmitter and receiver
solving a problem of parallel channel discrimination. The
transmitter is an arbitrary multipartite state $\rho$ which probes
the entire block by inputting $s$ systems per cell plus sending
additional $r$ systems directly to the receiver. The output state
$\rho_{\mathbf{x}^{i}}$ is detected by an optimal collective
measurement which provides the correct answer $\mathbf{x}^{i}$ up
to some error probability $P_{err}$. In the uncostrained readout,
$P_{err}$ goes to zero and Bob retrieves all the information
$H_{\max}$ from the block. If the readout is constrained, as in
the case of optical memories at fixed signal energy $n$, then
$P_{err}$ is nonzero. In this case, Bob retrieves a fraction of
the information $I\leq H_{\max}$ or, equivalently, he retrieves
all the information if Alice suitably increases the size of the
block while keeping $H_{\max}$ as constant.} \label{Fig3}
\end{figure}

The most general strategy of readout can be described as a problem of
\textquotedblleft parallel discrimination of quantum
channels\textquotedblright, where Bob probes the entire block in a parallel
fashion and detects the output via a collective quantum measurement. In order
to query the block, Bob uses $s$ signal systems per cell besides other
supplemental $r$ reference systems for the benefit of the output measurement.
The whole set of $ms+r$ systems is described by an arbitrary multipartite
state $\rho$ (see Fig.~\ref{Fig3}). At the output of the block, Bob has%
\begin{equation}
\rho_{\mathbf{x}^{i}}:=(\phi_{\mathbf{x}^{i}}^{\otimes s}\otimes I^{\otimes
r})(\rho)~,
\end{equation}
where the identity acts on the reference systems, while%
\begin{equation}
\phi_{\mathbf{x}^{i}}^{\otimes s}=\phi_{x_{1}^{i}}^{\otimes s}\otimes
\cdots\otimes\phi_{^{x_{m}^{i}}}^{\otimes s}%
\end{equation}
acts on the signal systems. This state is detected by a collective quantum
measurement, i.e., a general POVM with $l$ detection operators with outcome
$i$ corresponding to codeword $\mathbf{x}^{i}$. Clearly, the main goal for Bob
is to optimize both input state and output measurement in order to retrieve
the maximal information from the block.

It is intuitive to understand that, without constraints, Bob is always able to
retrieve all the information from the block, i.e.,
\begin{equation}
H_{\max}=-\sum_{i=0}^{l-1}p_{\mathbf{x}^{i}}\log p_{\mathbf{x}^{i}}%
\end{equation}
bits of information. However this is no longer the case if we impose
constraints on Bob's physical resources. As we know, when we consider optical
memories (bosonic setting), the optimization must be constrained in the input
energy, in particular, by fixing the mean total number of photons
$n$\ irradiated over each cell. In this case, if we consider low values of
$n$, the measurement will be affected by non-negligible error probability
$P_{err}$ and the information retrieved will be some value $I=I(n)$ between
$0$ and $H_{\max}$. In other words, data will be read with an average rate of
$R(n)=m^{-1}I(n)$ bits per cell.

It is important to note that, in the block-encoding model, an equivalent
approach consists of making the readout flawless by increasing the error
correction overhead in the block. In other words, for a given signal energy
$n$, we can determine the minimal size $m=m(n)$ of the block (and the
corresponding optimal classical code) which makes the error probability
$P_{err}$ negligible (i.e., reasonably close to zero~\cite{PirsSUPP}). In this
case, the readout is flawless and the block provides all the $H_{\max}$\ bits
of information. As a result, the rate now takes the form $R(n)=[m(n)]^{-1}%
H_{\max}$.

It is clear that, given an arbitrary block encoding $\{\phi_{\mathbf{x}^{i}%
},p_{\mathbf{x}^{i}}\}$ and an arbitrary multipartite transmitter which
irradiates $n$ mean photon per cell (see Fig.~\ref{Fig3}), the computation of
the rate $R(n)$\ is extremely difficult. However we can face the problem if we
consider transmitters which are separable with respect to the different cells
(more exactly, in tensor product form) and taking the limit for infinite block
($m\rightarrow\infty$). This allows us to introduce a simple description of
the memory (similar to the single-cell scenario) and, most importantly, to use
the Holevo bound as quantifier for the readable information, i.e., as
asymptotic rate $R(n)$. Then, the optimization of this rate over the
transmitters enables us to define the quantum reading capacity of the memory.

\section{Limit for infinite block: quantum reading capacity\label{InfiniteSEC}%
}

Digital memories typically store a great amount of data. This means that an
average\ memory is composed of a large number of encoding blocks. In
principle, we can also describe the memory as a single large block of cells
where Alice stores data by encoding a very long channel-codeword
$\phi_{\mathbf{x}^{i}}$\ chosen with some probability $p_{\mathbf{x}^{i}}$.
Considering the whole memory as a large encoding block allows us to
re-introduce a single-cell description. In fact, in the limit for
$m\rightarrow\infty$, each cell can be described (on overage) by a marginal
ensemble of quantum channels $\Phi=\left\{  \phi_{x},p_{x}\right\}  $ encoding
a corresponding marginal variable $X=\left\{  x,p_{x}\right\}  $. Thus,
independently from the actual classical code used to store information, the
description of a large classical memory can always be reduced to its marginal
cell, corresponding to a marginal ensemble of channels $\Phi=\left\{  \phi
_{x},p_{x}\right\}  $.

\begin{figure}[ptb]
\vspace{-0.4cm}
\par
\begin{center}
\includegraphics[width=0.5\textwidth] {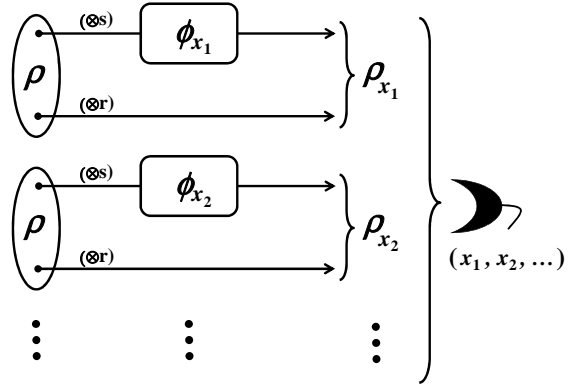}
\end{center}
\par
\vspace{-1.0cm}\caption{\textbf{Limit for infinite block.} A memory can be
described as a large (approximately infinite) encoding block, where each cell
encodes a marginal ensemble $\Phi=\{\phi_{x},p_{x}\}$. In order to read the
memory, Bob uses an multi-copy transmitter $\rho(s,r)^{\otimes\infty}%
=\rho(s,r)\otimes\rho(s,r)\otimes\cdots$, where each copy
$\rho(s,r)$ probes a different cell using $s$ signals generally
coupled with other $r$ references. All the outputs from the block
are collectively detected by an optimal quantum measurement which
reconstructs the asymptotic channel codeword.} \label{figINF}
\end{figure}

Despite this asymptotic simplification, the readout process is still too
difficult to be treated if we consider arbitrary multipartite states, i.e.,
generally entangled among different cells. Thus we restrict the readout to
input states which are tensor products. This means that Bob inputs an $\infty
$-copy state
\begin{equation}
\rho(s,r)^{\otimes\infty}=\rho(s,r)\otimes\rho(s,r)\otimes\cdots,
\end{equation}
where the single-copy $\rho(s,r)\in\mathcal{D}(\mathcal{H}_{S}^{\otimes
s}\otimes\mathcal{H}_{R}^{\otimes r})$ describes $s$ signal systems sent
through a target cell plus additional $r$ reference systems (see
Fig.~\ref{figINF}). Given the $\infty$-copy transmitter $\rho(s,r)^{\otimes
\infty}$ at the input of a memory with marginal cell $\Phi=\{\phi_{x},p_{x}%
\}$, the output is still in a tensor product form. The average output of each
cell is described by a marginal ensemble of states
\begin{equation}
\mathcal{E}=\{\rho_{x}(s,r),p_{x}\}~,
\end{equation}
where%
\begin{equation}
\rho_{x}(s,r)=(\phi_{x}^{\otimes s}\otimes I^{\otimes r})[\rho(s,r)]~.
\label{rhoxPARALLEL}%
\end{equation}
By applying an optimal collective measurement on all the outputs, the maximum
information per cell that can be retrieved is given by the Holevo bound%
\begin{equation}
\chi(\mathcal{E})=S\left[  \sum_{x}p_{x}\rho_{x}(s,r)\right]  -\sum_{x}%
p_{x}S\left[  \rho_{x}(s,r)\right]  \label{HolMemory}%
\end{equation}
where $S$ denotes the von Neumann entropy. The achievability of $\chi
(\mathcal{E})$\ is assured by the Holevo-Schumacher-Westmoreland
theorem~\cite{HSW}. Here it is important to note that the asymptotic readout
can be flawless. In other words, for a given marginal $\Phi=\{\phi_{x}%
,p_{x}\}$ there is always an asymptotic block code $\{\phi_{\mathbf{x}^{i}%
},p_{\mathbf{x}^{i}}\}$, giving that marginal, which allows the receiver to
give the correct answer $\mathbf{x}^{i}$ with asymptotically zero
error.

Given a memory with marginal cell $\Phi$, the Holevo information of
Eq.~(\ref{HolMemory}) depends on the input state $\rho(s,r)$ only. This means
that it can be represented as a \emph{conditional} Holevo information, that we
denote by $\chi\lbrack\Phi|\rho(s,r)]$. In other words $\chi\lbrack\Phi
|\rho(s,r)]$ represents the maximum information per cell which can be read
from a memory with marginal cell $\Phi$ if we use the transmitter $\rho(s,r)$.
The crucial task here is the optimization of $\chi\lbrack\Phi|\rho(s,r)]$ over
the transmitter. As a first step, we can consider the readout capacity for
fixed number of input systems $s$ and $r$, i.e.,%
\begin{equation}
C(\Phi|s,r)=\max_{\rho(s,r)}\chi\lbrack\Phi|\rho(s,r)]~. \label{C_sr}%
\end{equation}
Now, by optimizing Eq.~(\ref{C_sr}) over the number of input systems, we can
define the \textit{unconstrained} quantum reading capacity of the
memory~\cite{capp}%
\begin{equation}
C(\Phi)=\sup_{s,r}\max_{\rho(s,r)}\chi\lbrack\Phi|\rho(s,r)]~. \label{readCAP}%
\end{equation}
Since it is unconstrained, this capacity can be greatly simplified and
trivially computed. First of all the maximization can be reduced to pure
transmitters $\psi(s,r)$ as a simple consequence of the convexity of the
Holevo information~\cite{NC} (see Appendix~\ref{APPpure} for more details).
Then, also the use of the reference systems can be avoided. In other words, it
is sufficient to consider the unassisted capacity where we maximize over
$\psi(s,0)=\psi(s)$. Furthermore the supremum is achieved in the limit for
$s\rightarrow+\infty$, i.e., we can write
\begin{equation}
C(\Phi)=\lim_{s\rightarrow+\infty}\max_{\psi(s)}\chi\lbrack\Phi|\psi(s)]~.
\end{equation}
This quantity is the maximal possible since it equals the amount of
information stored in the marginal cell of the memory. This is given by the
Shannon entropy of the marginal variable $X=\left\{  x,p_{x}\right\}  $
encoded by the marginal ensemble $\Phi=\left\{  \phi_{x},p_{x}\right\}  $. In
other words, we have%
\begin{equation}
C(\Phi)=H(X)=-\sum_{x}p_{x}\log p_{x}~.
\end{equation}
The proof is trivial (see Appendix~\ref{APPtrivial} for details).

The notion of quantum reading capacity is non-trivial only in the presence of
physical constraints. This is what happens in the bosonic setting, where
optical memories are read by fixing the input signal energy. Thus, let us
consider an optical memory with marginal cell $\Phi=\{\phi_{x},p_{x}\}$ where
$\phi_{x}$\ represents a single-mode bosonic channel. As transmitter, now we
consider an $\infty$-copy state
\begin{equation}
\rho(s,r,n)^{\otimes\infty}=\rho(s,r,n)\otimes\rho(s,r,n)\otimes\cdots,
\end{equation}
where $\rho(s,r,n)\in\mathcal{D}(\mathcal{H}_{S}^{\otimes s}\otimes
\mathcal{H}_{R}^{\otimes r})$ describes $s$ signal modes, irradiating $n$ mean
photons on a target cell, plus\ additional $r$ reference modes bypassing the
cell. At the output we have an infinite tensor product of states of the form%
\begin{equation}
\rho_{x}(s,r,n)=(\phi_{x}^{\otimes s}\otimes I^{\otimes r})[\rho(s,r,n)]~,
\end{equation}
which are detected by an optimal collective measurement. In this way, Bob is
able to retrieve an average of $\chi\lbrack\Phi|\rho(s,r,n)]$ bits per cell.
Now, we must optimize this quantity over the input transmitters by taking the
signal energy $n$ fixed. This \textit{constrained} optimization leads to the
definition of the quantum reading capacity of the optical memory%
\begin{equation}
C(\Phi|n)=\sup_{s,r}\max_{\rho(s,r,n)}\chi\lbrack\Phi|\rho(s,r,n)]~.
\label{Cap_constrained}%
\end{equation}
This capacity represents the maximum information per cell which is readable
from an optical memory $\Phi$ by irradiating $n$ mean photons per cell. The
computation of Eq.~(\ref{Cap_constrained}) is not easy at all. As a matter of
fact we are only able to provide lower bounds by restricting the class of
transmitters involved in the maximization. We do not even know if the optimal
transmitters are pure or mixed.

Let us consider a set (or \textquotedblleft class\textquotedblright)
$\mathcal{P}$ of pure transmitters $\psi(s,r,n)$ which are characterized by
some general property which does not depend on $s$, $r$ and $n$ (for instance,
they could be constructed using states of particular kind, such as coherent
states). Then we can always construct the mixed-state transmitter%
\begin{equation}
\rho(s,r,n)=\int dy~p_{y}~\psi_{y}(s,r,n)~, \label{arbTRAproof}%
\end{equation}
where $~p_{y}\geqslant0,~\int dy~p_{y}=1$, and $\psi_{y}(s,r,n)\in\mathcal{P}%
$. Clearly the set of mixed-state transmitters identifies a larger class
$\mathcal{A}$ which includes $\mathcal{P}$. Now, we can define a lower-bound
to $C(\Phi|n)$ by optimizing over the class $\mathcal{A}$, i.e.,%
\begin{equation}
C(\Phi|n)\geq C_{\mathcal{A}}(\Phi|n)=\sup_{s,r}\max_{\rho(s,r,n)\in
\mathcal{A}}\chi\lbrack\Phi|\rho(s,r,n)]~.
\end{equation}
Similarly we can consider the further lower-bound
\begin{equation}
C_{\mathcal{A}}(\Phi|n)\geq C_{\mathcal{P}}(\Phi|n)=\sup_{s,r}\max
_{\psi(s,r,n)\in\mathcal{P}}\chi\lbrack\Phi|\psi(s,r,n)]~. \label{CaCp}%
\end{equation}
Here we first ask: is there some class $\mathcal{P}$\ that allows to put an
equality in Eq.~(\ref{CaCp}), i.e., $C_{\mathcal{P}}(\Phi|n)=C_{\mathcal{A}%
}(\Phi|n)$? Then, is it possible to extend this class to all the pure
transmitters, so that $C_{\mathcal{P}}(\Phi|n)=C(\Phi|n)$?

Unfortunately we are not able to answer the second question, so that the issue
of the purity of the optimal transmitters remains unsolved. However, we are
able to find classes for which $C_{\mathcal{P}}(\Phi|n)=C_{\mathcal{A}}%
(\Phi|n)$. For this sake, a sufficient criterion is the concavity of
$C_{\mathcal{P}}(\Phi|n)$.

\begin{lemma}
\label{OPS}If $C_{\mathcal{P}}(\Phi|n)$ is concave in $n$, then we have%
\begin{equation}
C_{\mathcal{P}}(\Phi|n)=C_{\mathcal{A}}(\Phi|n)~. \label{Cequal}%
\end{equation}

\end{lemma}

\noindent\textbf{Proof.} Let us consider the transmitter of
Eq.~(\ref{arbTRAproof}), whose signal energy (mean number of photons) can be
written as%
\begin{equation}
n=\int dy~p_{y}~n_{y}\,,~n_{y}=\langle\psi_{y}|\hat{n}|\psi_{y}\rangle~.
\end{equation}
Given this transmitter at the input of a marginal cell $\Phi$, we can bound
the conditional Holevo information%
\begin{align}
\chi\lbrack\Phi|\rho(s,r,n)]  &  \leqslant\int dy~p_{y}~\chi\lbrack\Phi
|\psi_{y}]\label{p1}\\
&  \leqslant\int dy~p_{y}~C_{\mathcal{P}}(\Phi|n_{y})\label{p2}\\
&  \leqslant C_{\mathcal{P}}\left(  \Phi|\int dy~p_{y}~n_{y}\right)
\label{p3b}\\
&  =C_{\mathcal{P}}\left(  \Phi|n\right)  ~, \label{p4}%
\end{align}
where we have used the convexity of $\chi$ in the first inequality~(\ref{p1}),
the definition of $C_{\mathcal{P}}(\Phi|n)$ in the second inequality~(\ref{p2}%
) and its concavity in the last inequality~(\ref{p3b}). It is clear that
Eqs.~(\ref{p1})-(\ref{p4}) hold for every $\rho(s,r,n)\in\mathcal{A}$ and
every $s$ and $r$. As a result, we can write%
\begin{equation}
\sup_{s,r}\max_{\rho(s,r,n)\in\mathcal{A}}\chi\lbrack\Phi|\rho
(s,r,n)]=C_{\mathcal{A}}(\Phi|n)\leqslant C_{\mathcal{P}}\left(
\Phi|n\right)  ~,
\end{equation}
which, combined with Eq.~(\ref{CaCp}), gives the result of Eq.~(\ref{Cequal}%
).$~\blacksquare$

In the following section we show that an important class $\mathcal{P}$\ for
which $C_{\mathcal{P}}(\Phi|n)$ is concave is the one of the coherent-state
transmitters. This means that $C_{\mathcal{P}}(\Phi|n)=C_{\mathcal{A}}%
(\Phi|n)$, where $\mathcal{A}$ is the class of the classical transmitters
(constructed by convex combination via the $P$-function). Thanks to this
result we can compute an analytical bound for the readout performance of all
the classical transmitters, that we call \textquotedblleft classical reading
capacity\textquotedblright. This capacity represents the multi-cell
generalization of the classical discrimination bound of
Sec.~\ref{SimpleMemorySEC} and provides a simple lower bound to the quantum
reading capacity. In Sec.~\ref{BoundSEC}\ we compute its analytical formula
for the most basic optical memories. Then, as we will show in
Sec.~\ref{NonclassicalSEC}, this classical bound can be easily outperformed by
nonclassical transmitters, thus proving its separation from the quantum
reading capacity.

\section{Classical reading capacity\label{BoundSEC}}

Let us consider an optical memory which is the multi-cell generalization of
the binary model described in Sec.~\ref{SimpleMemorySEC}. In the single-cell
model of Sec.~\ref{SimpleMemorySEC}, information was written in each cell in
an independent fashion, by encoding one of two possible pure-loss channels,
$\phi_{0}$ and $\phi_{1}$ (binary cell). Here we consider the multi-cell
version, where Alice stores a channel codeword in the whole optical memory
regarded as an infinite block. In particular, the block encoding is such that
the marginal cell is described by a binary ensemble%
\begin{equation}
\tilde{\Phi}=\{\phi_{0},p,\phi_{1},1-p\}
\end{equation}
where $0\leq p\leq1$ and $\phi_{u}$ is a pure-loss channel with transmission
$\kappa_{u}$. Alternatively, we can use the notation
\begin{equation}
\tilde{\Phi}=\{\kappa_{0},p,\kappa_{1},1-p\}~.
\end{equation}
Given this kind of memory, we consider the input%
\begin{equation}
\rho_{c}(s,r,n)^{\otimes\infty}=\rho_{c}(s,r,n)\otimes\rho_{c}(s,r,n)\otimes
\cdots,
\end{equation}
where $\rho_{c}(s,r,n)$ is an arbitrary classical transmitter with $s$
signals, $r$ references and $n$ mean photons. The average information which
can be read from each cell is provided by the Holevo quantity $\chi
\lbrack\tilde{\Phi}|\rho_{c}(s,r,n)]$. Now, by optimizing over the classical
transmitters we can define the lower-bound
\begin{equation}
C(\tilde{\Phi}|n)\geq C_{c}(\tilde{\Phi}|n)=\sup_{s,r}\max_{\rho_{c}%
(s,r,n)}\chi\lbrack\tilde{\Phi}|\rho_{c}(s,r,n)]~, \label{CDB_multi}%
\end{equation}
which represents the classical reading capacity of the optical memory
$\tilde{\Phi}$. This capacity represents the multi-cell version of the
classical discrimination bound of Sec.~\ref{SimpleMemorySEC}. As before, we
can provide a simple analytical result.

\begin{theorem}
Let us consider an optical memory with binary marginal cell $\tilde{\Phi
}=\{\kappa_{0},p,\kappa_{1},1-p\}$ which is read by a classical transmitter
signalling $n$\ mean photons. Then, the maximum information per cell which can
be read is asymptotically equal to%
\begin{equation}
C_{c}(\tilde{\Phi}|n)=H(\xi)~, \label{mainTHEO}%
\end{equation}
where $H$ is the binary Shannon entropy and%
\begin{equation}
\xi=\frac{1}{2}+\frac{1}{2}\sqrt{1-4p(1-p)\left[  1-e^{-n(\sqrt{\kappa_{1}%
}-\sqrt{\kappa_{0}})^{2}}\right]  }~. \label{csiTHEO}%
\end{equation}
In particular, the bound $C_{c}(\tilde{\Phi}|n)$ can be reached by using a
coherent-state transmitter $\rho_{coh}(1,0,n)=|\sqrt{n}\rangle_{S}\langle
\sqrt{n}|$, i.e., a single-mode coherent state with $n$ mean photons.
\end{theorem}

\noindent\textbf{Proof.}~Let us consider the class $\mathcal{P}=coh$ of
coherent-state transmitters $\rho_{coh}(s,r,n)$. By convex combination we
construct the class $\mathcal{A}=c$\ of the classical transmitters $\rho
_{c}(s,r,n)$. The first step of the proof is the computation of $C_{coh}%
(\tilde{\Phi}|n)$, i.e., the readout capacity restricted to coherent state
transmitters. We first prove that $C_{coh}(\tilde{\Phi}|n)=\chi\lbrack
\tilde{\Phi}|\rho_{coh}(1,0,n)]$, i.e., the optimal coherent-state transmitter
is the single-mode coherent state $|\sqrt{n}\rangle_{S}\langle\sqrt{n}|$.
Then, we analytically compute $\chi\lbrack\tilde{\Phi}|\rho_{coh}(1,0,n)]$.
Since this quantity turns out to be concave in $n$, we can use Lemma~\ref{OPS}
and state $C_{coh}(\tilde{\Phi}|n)=C_{c}(\tilde{\Phi}|n)$, thus achieving the
result of the theorem.

Given a coherent-state transmitter
\begin{align}
\rho_{coh}(s,r,n)  &  =\sigma(\alpha)\otimes\gamma(\beta)\nonumber\\
&  =\bigotimes_{i=1}^{s}|\alpha_{i}\rangle_{S}\langle\alpha_{i}|\otimes
\bigotimes_{i=1}^{r}|\beta_{i}\rangle_{R}\langle\beta_{i}|~ \label{input1}%
\end{align}
at the input of the cell $\tilde{\Phi}$, we have the output
\begin{align}
\rho_{u}  &  =\phi_{u}^{\otimes s}[\sigma(\alpha)]\otimes\gamma(\beta
)\nonumber\\
&  =\bigotimes_{i=1}^{s}\phi_{u}(|\alpha_{i}\rangle_{S}\langle\alpha
_{i}|)\otimes\bigotimes_{i=1}^{r}|\beta_{i}\rangle_{R}\langle\beta
_{i}|\nonumber\\
&  =\bigotimes_{i=1}^{s}|\sqrt{\kappa_{u}}\alpha_{i}\rangle_{S}\langle
\sqrt{\kappa_{u}}\alpha_{i}|\otimes\bigotimes_{i=1}^{r}|\beta_{i}\rangle
_{R}\langle\beta_{i}|~, \label{output1}%
\end{align}
which is still a multimode coherent state. This is a simple consequence of the
fact that $\phi_{0}$ and $\phi_{1}$ are pure-loss channels. Since we are
computing the Holevo information on the output ensemble, we have the freedom
to apply a unitary transformation over $\rho_{u}$. By using a suitable
sequence of beam splitters and phase-shifters we can always transform
$\rho_{u}$ into the state%
\begin{equation}
|\sqrt{\kappa_{u}n}\rangle_{S}\langle\sqrt{\kappa_{u}n}|\otimes|0\rangle
\langle0|^{\otimes r+s-1}~.
\end{equation}
Then, since the Holevo information does not change under the adding of
systems, we can trace the $r+s-1$ vacua and just consider the single-mode
output state%
\begin{equation}
\rho_{u}=|\sqrt{\kappa_{u}n}\rangle_{S}\langle\sqrt{\kappa_{u}n}|~.
\label{output2}%
\end{equation}
This can be achieved by considering a single-mode coherent state transmitter
\begin{equation}
\rho_{coh}(1,0,n)=|\sqrt{n}\rangle_{S}\langle\sqrt{n}| \label{input2}%
\end{equation}
at the input of the pure-loss channel $\phi_{u}$. For fixed marginal cell
$\tilde{\Phi}$ and fixed input energy $n$, the reduction from the multimode
input of Eq.~(\ref{input1}) to the single-mode output of Eq.~(\ref{input2}) is
always possible, independently from the actual number of systems, $s$ and $r$,
and the specific form of the transmitter $\rho_{coh}(s,r,n)$. Thus, we can
write
\begin{align}
C_{coh}(\tilde{\Phi}|n)  &  =\sup_{s,r}\max_{\rho_{coh}(s,r,n)}\chi
\lbrack\tilde{\Phi}|\rho_{coh}(s,r,n)]\nonumber\\
&  =\chi\lbrack\tilde{\Phi}|\rho_{coh}(1,0,n)]~.
\end{align}
In other words, the optimal coherent-state transmitter is the single-mode
coherent state $|\sqrt{n}\rangle_{S}\langle\sqrt{n}|$. The next step is the
analytical computation of $\chi\lbrack\tilde{\Phi}|\rho_{coh}(1,0,n)]$. After
some Algebra we get%
\[
\chi\lbrack\tilde{\Phi}|\rho_{coh}(1,0,n)]=H(\xi)~,
\]
where $H$ is the binary formula of the Shannon entropy and $\xi=\xi(\kappa
_{0},\kappa_{1},p,n)$ is given in Eq.~(\ref{csiTHEO}). One can easily check
that $H(\xi)$ is a concave function of $n$, for any $\kappa_{0}$, $\kappa_{1}%
$, and $p$. Since $C_{coh}(\tilde{\Phi}|n)=H(\xi)$ is concave in the energy
$n$, we can apply Lemma~\ref{OPS} by setting $\mathcal{P}=coh$ and
$\mathcal{A}=c$. Thus we get $C_{c}(\tilde{\Phi}|n)=C_{coh}(\tilde{\Phi
}|n)=H(\xi)$ which is the result of Eq.~(\ref{mainTHEO}). It is clear that the
optimal classical transmitter coincides with the optimal coherent-state
transmitter which is given by $\rho_{coh}(1,0,n)$.~$\blacksquare$

\begin{figure}[ptbh]
\vspace{+0.1cm}
\par
\begin{center}
\includegraphics[width=0.37\textwidth] {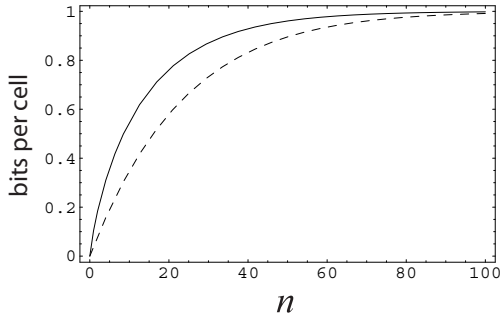}
\end{center}
\par
\vspace{-0.5cm} \caption{Maximum number of bits per cell read by classical
transmitters as a function of the signal energy $n$ (mean number of photons).
We compare the two classical discrimination bounds: $C_{c}(\bar{\Phi}|n)$
\ (multi-cell readout, solid line) and $I_{c}(\bar{\Phi}|n)$ (single-cell
readout, dashed line). We consider a memory with binary marginal cell
$\bar{\Phi}=\{\kappa_{0},\kappa_{1}\}$ where $\kappa_{0}=0.5$ and $\kappa
_{1}=0.9$.}%
\label{classBOUNDpic}%
\end{figure}

It is interesting to compare the single-cell and multi-cell classical
discrimination bounds, in order to estimate the gain which is provided by the
parallel readout of the cells. For a direct comparison, let us set $p=1/2$, so
that the binary cell $\tilde{\Phi}$ is described by $\bar{\Phi}=\{\kappa
_{0},1/2,\kappa_{1},1/2\}=\{\kappa_{0},\kappa_{1}\}$. Then, we compare the
maximum information which achievable by using classical transmitters in the
multi-cell readout, i.e., the classical reading capacity $C_{c}(\bar{\Phi}%
|n)$, with the maximum information which is achievable by classical
transmitters in the single-cell readout, i.e., the classical discrimination
bound $I_{c}(\bar{\Phi}|n)$ given in Eq.~(\ref{classBOUNDinfo}). As shown in
Fig.~\ref{classBOUNDpic} the advantage is quite evident.

In the following Sec.~\ref{NonclassicalSEC}, we will construct examples of
nonclassical transmitters which are able to outperform the classical reading
capacity. This will prove the separation between the quantum reading and the
classical reading capacities, thus showing the advantages of quantum reading
in the multi-cell scenario.

\section{Nonclassical transmitters\label{NonclassicalSEC}}

As before, let us consider an optical memory with binary marginal cell
$\bar{\Phi}=\{\kappa_{0},\kappa_{1}\}$. This time we assume that it is read by
using a nonclassical transmitter $\rho_{nc}(s,r,n)$. Since we are in the
asymptotic multi-cell scenario, we clearly assume an $\infty$-copy input
$\rho_{nc}(s,r,n)\otimes\rho_{nc}(s,r,n)\otimes\cdots$ together with an
optimal collective measurement of the output. The maximum number of readable
bits per cell is given by the conditional Holevo information $\chi\lbrack
\bar{\Phi}|\rho_{nc}(s,r,n)]$. Now we ask: is this quantity bigger than the
classical reading capacity $C_{c}(\bar{\Phi}|n)$?

\begin{figure}[ptbh]
\vspace{-0.0cm}
\par
\begin{center}
\includegraphics[width=0.48\textwidth]{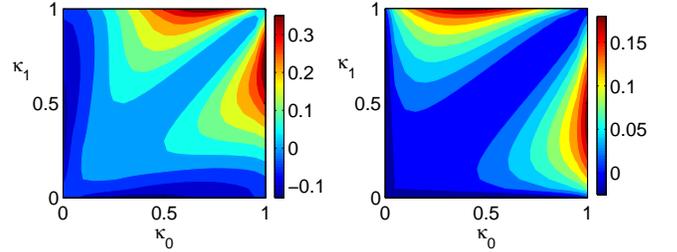}
\end{center}
\par
\vspace{-0.7cm}\caption{(Color online) Information gain $G$ versus
reflectivities, $\kappa_{0}$ and $\kappa_{1}$, for $n=5$ (left panel) and
$n=1$ (right panel). Here $G$ provides the number of bits per cell which are
gained by the single-copy EPR\ transmitter $|\xi\rangle\left\langle
\xi\right\vert $ over all the classical transmitters in the readout of an
optical memory with marginal cell $\bar{\Phi}=\{\kappa_{0},\kappa_{1}\}$. Note
that the highest values of $G$ occur for $\kappa_{0}$ or $\kappa_{1}$ close to
$1$ (high reflectivities).}%
\label{gainPIC}%
\end{figure}

The first design of nonclassical transmitter is the EPR\ transmitter
$\rho_{epr}(s,s,n)=|\xi\rangle\left\langle \xi\right\vert ^{\otimes s}$\ which
has been first discussed in Sec.~\ref{SimpleMemorySEC}. In order to beat
classical transmitters, it is sufficient to consider $\rho_{epr}%
(1,1,n)=|\xi\rangle\left\langle \xi\right\vert $, i.e., a single TMSV\ state
per cell. This means that we have one signal mode $S$, irradiating $n$ mean
photons over a target cell, which is entangled with one reference mode $R$. To
quantify the advantage we consider the information gain
\[
G=\chi\lbrack\bar{\Phi}|\rho_{epr}(1,1,n)]-C_{c}(\bar{\Phi}|n)
\]
and check its positivity. If $G>0$ then the EPR\ transmitter $\rho
_{epr}(1,1,n)$ beats all the classical transmitters, retrieving
$G$ bits per cell more than any classical strategy. As shown in
Fig.~\ref{gainPIC}, we have $G>0$ in the regime of low photons and
high reflectivities (i.e., $\kappa_{0}$ or $\kappa_{1}$ close to
$1$). This is the typical regime where the quantum reading of
optical memories is advantageous, as also investigated in the
single-cell scenario~\cite{Pirs}.

\begin{figure}[tbh]
\vspace{-0.3cm}
\par
\begin{center}
\includegraphics[width=0.5\textwidth]{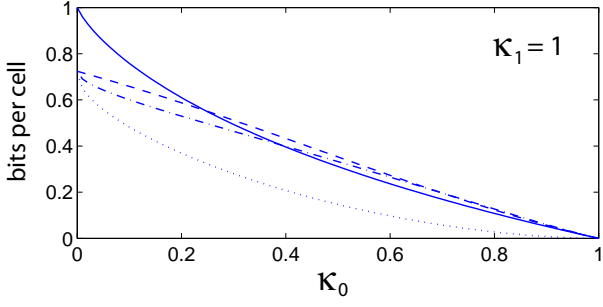}
\end{center}
\par
\vspace{-0.6cm} \caption{Number of bits per cell as a function of $\kappa_{0}%
$, for $\kappa_{1}=1$ (ideal memories) and $n$ $=1$ mean photons per cell. We
compare the classical reading capacity (dotted line) with the Holevo
information retrieved by various nonclassical transmitters: EPR transmitter
(dash-dotted line), NOON state transmitter (dashed line) and Fock state
transmitter (solid line).}%
\label{kappa1}%
\end{figure}

As evident from Fig.~\ref{gainPIC} the best situation corresponds to having
one of the two reflectivities equal to $1$, i.e., for an \textquotedblleft
ideal memory\textquotedblright\ $\bar{\Phi}=\{\kappa_{0}<\kappa_{1},\kappa
_{1}=1\}$. Given such a memory, we explicitly compare the information read by
an EPR\ transmitter $\chi_{epr}=\chi\lbrack\bar{\Phi}|\rho_{epr}(1,1,n)]$ with
the classical reading capacity $C_{c}(\bar{\Phi}|n)$ at low signal energy. As
shown in Fig.~\ref{kappa1} for $n=1$, the EPR\ transmitter is always able to
beat the classical bound.

It is important to note that we can construct other simple examples of
nonclassical transmitters which can outperform the classical reading capacity.
An alternative example of nonclassical transmitter can be taken again of the
form $\rho_{nc}(1,1,n)$ and corresponds to the NOON state~\cite{NOON1,NOON2}%
\begin{equation}
|NOON\rangle=2^{-1/2}(|2n\rangle_{S}|0\rangle_{R}+|0\rangle_{S}|2n\rangle
_{R})~,
\end{equation}
where signal and reference are again entangled. A further example of
nonclassical transmitter is of the form $\rho_{nc}(1,0,n)$, i.e., not
involving the reference mode. This is the Fock state%
\begin{equation}
|n\rangle_{S}=(n!)^{-1/2}(a_{S}^{\dag})^{n}|0\rangle_{S}~.
\end{equation}
As shown in Fig.~\ref{kappa1}, these transmitters can beat not
only the classical reading capacity but also the EPR\ transmitter
$|\xi\rangle \left\langle \xi\right\vert $ for low values of
$\kappa_{0}$. Recently, these kinds of transmitters have been also
studied by Ref.~\cite{NairLAST}\ in the basic context of quantum
reading with single-cell readout.

\begin{figure}[tbh]
\vspace{-0.3cm}
\par
\begin{center}
\includegraphics[width=0.5\textwidth]{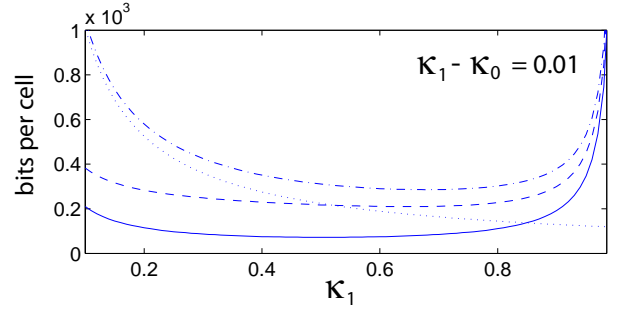}
\end{center}
\par
\vspace{-0.6cm} \caption{Number of bits per cell as a function of $\kappa_{1}%
$, for $\kappa_{1}-\kappa_{0}=0.01$ (close reflectivities) and $n=1$. We
compare the classical reading capacity (dotted line) with different
nonclassical transmitters: EPR transmitter (dash-dotted line), NOON state
transmitter (dashed line) and Fock state transmitter (solid line).}%
\label{kappad01}%
\end{figure}

It is interesting to compare the performances of all these transmitters in the
low-energy readout of optical memories with very close reflectivities. This is
shown in Fig.~\ref{kappad01} for $\kappa_{1}-\kappa_{0}=0.01$ and $n=1$. The
EPR\ transmitter $|\xi\rangle\left\langle \xi\right\vert $ is optimal almost
everywhere, while the classical bound beats the other nonclassical
transmitters for low values of $\kappa_{1}$ (and $\kappa_{0}$). This is also
compatible with the result of optimality of the TMSV\ state for the problem of
estimating the unknown loss parameter of a bosonic channel \cite{estimo}. As
evident from Fig.~\ref{kappad01} the bigger separation from the classical
bound occurs for high reflectivities, i.e., $\kappa_{1}$ close to
$1$.

\begin{figure}[tbh] \vspace{-0.3cm}
\par
\begin{center}
\includegraphics[width=0.5\textwidth]{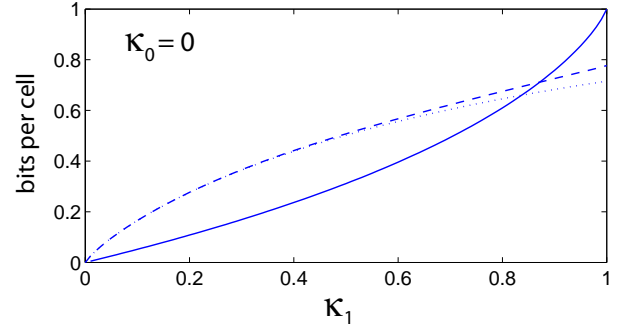}
\end{center}
\par
\vspace{-0.6cm} \caption{Number of bits per cell as a function of $\kappa_{1}%
$, for $\kappa_{0}=0$ and $n=1$. We compare the classical reading capacity
(dotted line) with the squeezed coherent state transmitter (dashed line) and
the Fock state transmitter (solid line).}%
\label{kappa0}%
\end{figure}

It is also interesting to see what happens in the regime of low reflectivity
by considering a binary marginal cell $\bar{\Phi}=\{\kappa_{0},\kappa_{1}%
\}$\ with $\kappa_{0}=0$. For this comparison we introduce another
nonclassical transmitter of the form $\rho_{nc}(1,0,n)$. This is
the squeezed coherent state
$|\alpha,\xi\rangle=D(\alpha)S(\xi)|0\rangle$, where $D(\alpha)$
is the displacement operator and $S(\xi)$ the squeezing
operator~\cite{Walls}. The squeezed coherent state is chosen with
the squeezing orthogonal to the displacement direction. Then we
choose two real parameters, $\alpha$ and $\xi$, which are
optimized under the condition $\alpha^{2}+\sinh^{2}{\xi}=n$,
imposed by the mean photon-number constraint. As shown in
Fig.~\ref{kappa0}, the presence of squeezing is sufficient to
outperform the classical reading capacity in the regime of low
reflectivity. However, better performances can be achieved by the
Fock state considering high values of
$\kappa_{1}$.

From the previous analysis it is evident that, in the regime of low photon
number (down to one photon per cell), we can easily find nonclassical
transmitters able to beat any classical transmitters, i.e., the classical
reading capacity. This is particularly evident for high reflectivities
($\kappa_{1}$ or $\kappa_{0}$ close to $1$). Thus, for the most basic optical
memories, classical and quantum reading capacities are separated at low
energies. In other words, the advantages of quantum reading are fully extended
from the single- to the optimal multi-cell scenario.

\begin{figure}[tbh]
\vspace{-0.3cm}
\par
\begin{center}
\includegraphics[width=0.5\textwidth]{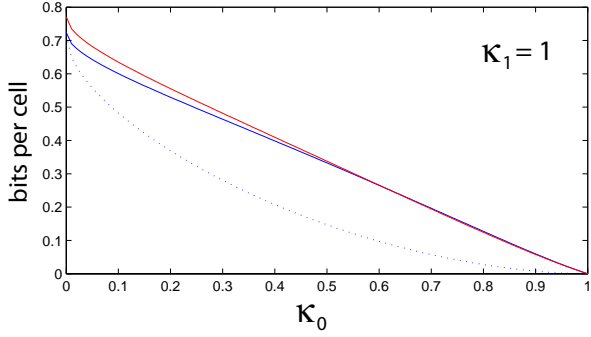}
\end{center}
\par
\vspace{-0.6cm}\caption{Number of bits per cell as a function of $\kappa_{0}$,
for $\kappa_{1}=1$ (ideal memory) and $n=1$. We compare the classical reading
capacity (dotted line) with two different EPR transmitters: $\rho
_{epr}(1,1,n)$ (lower solid line) and $\rho_{epr}(2,2,n)$ (upper solid line).}%
\label{super}%
\end{figure}

At this point a series of important considerations are in order. First of all,
note that we have only considered nonclassical transmitters irradiating one
signal per mode (entangled or not with a single reference mode), i.e.,
transmitters of the kind $\rho_{nc}(1,0,n)$ or $\rho_{nc}(1,1,n)$. The reason
is because these transmitters are sufficient to beat the classical bound.
However, better performances can be reached by optimizing over the number
signals and references. In the case of the EPR\ transmitters, we expect that
$\rho_{epr}(2,2,n)$, which is composed of two TMSV states signalling $n/2$
mean photons each, is able to outperform $\rho_{epr}(1,1,n)$, i.e., a single
TMSV state signalling $n$ mean photons. This is shown in Fig.~\ref{super} for
the case of an ideal memory and $n=1$ mean photons. This advantage could
further improve for EPR\ transmitters $\rho_{epr}(s,s,n)$ with higher values
of $s$. For this reason, in order to reach the quantum reading capacity, it is
necessary to optimize over an arbitrary number of signal and reference modes,
as foreseen by the general definition of Eq.~(\ref{Cap_constrained}%
).

\begin{figure}[tbh]
\vspace{+0.0cm}
\par
\begin{center}
\includegraphics[width=0.35\textwidth]{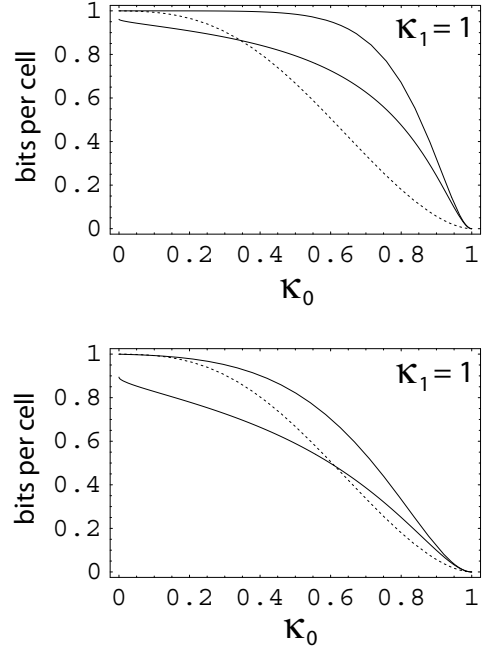}
\end{center}
\par
\vspace{-0.3cm}\caption{\textbf{Upper panel.} Number of bits per cell as a
function of $\kappa_{0}$, for $\kappa_{1}=1$ (ideal memory) and $n=5$. We
compare the classical reading capacity $C_{c}(\bar{\Phi}|n)$ (multi-cell
readout, dotted line) with the EPR transmitters used in the single-cell
readout (solid lines). The lower solid line refers to $Q(\bar{\Phi}|1,n)$,
i.e., a single energetic TMSV state $\rho_{epr}(1,1,n)$, while the upper solid
line refers to $Q(\bar{\Phi}|\infty,n)$, i.e., the optimal EPR transmitter
$\rho_{epr}(\infty,\infty,n)$ corresponding to infinite copies of TMSV\ states
with vanishing signal energy. \textbf{Lower panel. }As in the upper panel,
except that now we compare $C_{c}(\bar{\Phi}|n)$ with $Q(\bar{\Phi}|1,n/2)$
and $Q(\bar{\Phi}|\infty,n/2)$. Despite we assume a stronger energy constraint
involving the mean total number of photons in both signal and reference modes,
the single-cell quantum reading is still able to outperform the asymptotic
multi-cell classical reading.}%
\label{crossTOT}%
\end{figure}

Another important consideration is related to the practical realization of
quantum reading. In order to be experimentally feasible, the detection scheme
should be as simple as possible. For this reason, it is interesting to compare
the classical reading capacity (which refers to the general multi-cell
readout) with the performances of EPR\ transmitters in the single-cell
scenario, where each cell is detected independently from the others. Thus, we
consider an ideal memory $\bar{\Phi}=\{\kappa_{0}<\kappa_{1},\kappa_{1}=1\}$
which is irradiated by a few mean photons per cell (in particular, we can
consider $n=5$). Given this memory, we compare the optimal performance
$C_{c}(\bar{\Phi}|n)$ of classical transmitters assuming the multi-cell
readout (asymptotic collective measurement) with the performance of EPR
transmitters $\rho_{epr}(s,s,n)$ assuming the single-cell readout (individual
cell-by-cell measurements). The latter quantity is given by the mutual
information
\begin{equation}
I_{epr}(\bar{\Phi}|s,n)=1-H\{P[\bar{\Phi}|\rho_{epr}(s,s,n)]\},
\end{equation}
where $H$ is the binary formula for the Shannon entropy and $P[\bar{\Phi}%
|\rho_{epr}(s,s,n)]$ is the error probability of the single-cell
readout. One can compute the upper bound
\begin{equation}
P[\bar{\Phi}|\rho_{epr}(s,s,n)]\leq\Theta:=\frac{1}{2}\left[  1+\frac{n}%
{s}(1-\sqrt{\kappa_{0}})\right]  ^{-2s}~, \label{thetaFIN}%
\end{equation}
which provides a lower bound for the mutual information%
\begin{equation}
I_{epr}(\bar{\Phi}|s,n)\geq Q(\bar{\Phi}|s,n):=1-H(\Theta)~.
\end{equation}
Thus, $Q(\bar{\Phi}|s,n)$ provides the \textit{minimum} number of bits per
cell which are read by an EPR\ transmitter $\rho_{epr}(s,s,n)$. For fixed
signal energy $n$, it is trivial to check that this quantity is increasing in
$s$. This means that for any integer $s$ we have%
\begin{equation}
Q(\bar{\Phi}|1,n)\leq Q(\bar{\Phi}|s,n)\leq Q(\bar{\Phi}|\infty,n)~,
\end{equation}
where $Q(\bar{\Phi}|1,n)$ corresponds to a single energetic TMSV state
$\rho_{epr}(1,1,n)$ while $Q(\bar{\Phi}|\infty,n)$\ corresponds to $\rho
_{epr}(\infty,\infty,n)$, i.e., infinite copies of TMSV states with vanishing
energy. The quantity $Q(\bar{\Phi}|\infty,n)$ is computed by taking the limit
for $s\rightarrow\infty$\ in Eq.~(\ref{thetaFIN}). In this limit, we have
$\Theta\rightarrow\theta$, with $\theta$ given in Eq.~(\ref{thetaINI}). In the
upper panel of Fig.~\ref{crossTOT} we explicitly compare the two extremal
values $Q(\bar{\Phi}|1,n)$ and $Q(\bar{\Phi}|\infty,n)$ with the classical
reading capacity $C_{c}(\bar{\Phi}|n)$. As we can see, the single-cell quantum
reading is able to beat the asymptotic multi-cell classical
reading.

Finally, it is interesting to check if the single-cell quantum reading
represents a superior readout strategy even if we consider a stronger energy
constraint, for instance if we fix the mean total number of photons in both
the signal and reference modes for each copy of the transmitter. Note that
this approach has been first considered in Ref.~\cite{Carmen} for
individuating the optimal thermal probes (i.e., the optimal squeezed thermal
vacua) for detecting the presence of loss in bosonic channels. While this
stronger energy constraint does not make any difference for the classical
reading capacity (since the optimal classical transmitter involves signal
modes only) it clearly affects the EPR\ transmitters where the mean total
energy of the TMSV states is split exactly in two between signal and reference
modes. Thus, imposing this stronger energy constraint corresponds to compare
$C_{c}(\bar{\Phi}|n)$ with $Q(\bar{\Phi}|1,n/2)$ and $Q(\bar{\Phi}%
|\infty,n/2)$. As shown by the lower panel of Fig.~\ref{crossTOT}, we see that
the single-cell quantum reading is still able to beat the asymptotic
multi-cell classical reading.

\section{Conclusion\label{SEC_Conclusion}}

In this paper we have extended the basic model of quantum reading to the
optimal and asymptotic multi-cell scenario. Here the classical memory is
modelled as a large block of cells where information is stored by encoding a
suitable channel codeword (channel encoding). This information is then
retrieved by probing the whole memory in a parallel fashion and detecting the
output via an optimal collective measurement (channel discrimination). In this
general scenario, we define the quantum reading capacity of the memory which
is a nontrivial quantity to compute under the assumption of physical
constraints for the decoder. In the case of optical memories, where data
encoding is realized by bosonic channels, the main physical constraint is
energetic. This leads to define the quantum reading capacity of an optical
memory as the maximum number of bits per cell which can be read by irradiating
$n$ mean photons per cell.

Despite the general calculation of this capacity is extremely difficult, we
are still able to provide non-trivial lower bounds in the case of optical
memories with binary cells. The first lower bound, which we have called
classical reading capacity, represents the maximum number of bits per cell
which can be read by classical transmitters. This bound has a remarkably
simple analytical formula and can be achieved by using a single-mode coherent
state transmitter. Besides this result, we have also computed other bounds by
considering particular kinds of nonclassical transmitters, including the ones
constructed with TMSV states (EPR\ transmitters), NOON states and Fock states.
Then we have shown that, in the regime of few photons and high reflectivities,
these nonclassical transmitters are able to outperform any classical
transmitter, thus showing the separation between the classical and the quantum
reading capacities. It is remarkable that using a single-mode or two-mode
transmitter per cell is already sufficient to beat any classical strategy.
Furthermore the classical reading capacity can be outperformed even if we
restrict the EPR\ transmitters to the single-cell readout and we adopt the
stronger energy constraint where the energy of the reference modes is also
taken into account.

In conclusion, our study considers the optimal multi-cell encoding for
classical memories where we fully extends the advantages of quantum reading,
i.e., the readout by nonclassical transmitters. These advantages are
particularly evident in the regime of few photons with nontrivial consequences
for the technology of data storage.

\section{Acknowledgments}

The research leading to these results has received funding from the EU under
grant agreement No.~MOIF-CT-2006-039703 and the Italian Ministry of University
and Research under the FIRB-IDEAS project RBID08B3FM. The work of C. L. and S.
M. has been supported by EU under the FET-Open grant agreement HIP, No.
FP7-ICT-221889. S.P. would like to thank Jeffrey H. Shapiro, Saikat Guha and
Ranjith Nair for discussions.

\appendix

\section{Miscellaneous proofs}

\subsection{Reduction to pure transmitters\label{APPpure}}

For the sake of completeness, we show here that the maximization in
Eq.~(\ref{readCAP}) can be restricted to pure transmitters. This is a trivial
consequence of the convexity of the Holevo information.

Let us consider a classical memory with marginal cell $\Phi$ which is read by
an arbitrary transmitter with $s$ signals and $r$ references%
\begin{equation}
\rho(s,r)=\int dy~p_{y}~\psi_{y}~,~~~~\psi_{y}=|\psi_{y}\rangle\langle\psi
_{y}|~,
\end{equation}
where $p_{y}\geqslant0$ and $\int dy~p_{y}=1$. Then, the conditional Holevo
information obeys the inequality
\begin{equation}
\chi\lbrack\Phi|\rho(s,r)]\leqslant\int dy~p_{y}~\chi(\Phi|\psi_{y})~.
\label{toprove}%
\end{equation}
In order to prove Eq.~(\ref{toprove}), let us consider an auxiliary system
associated with the variable $y$. We denote by $\{|y\rangle\}$ an orthonormal
basis of this system. Then, we can express the transmitter as%
\begin{equation}
\rho(s,r)=\mathrm{Tr}_{y}\left(  \int dy~p_{y}~\psi_{y}\otimes|y\rangle\langle
y|\right)  ~.
\end{equation}
Since the Holevo information cannot increase under partial trace, then we
have
\begin{align}
\chi\lbrack\Phi|\rho(s,r)]  &  \leqslant\chi\left(  \Phi|\int dy~p_{y}%
~\psi_{y}\otimes|y\rangle\langle y|\right) \\
&  =\int dy~p_{y}~\chi\left(  \Phi|\psi_{y}\right)  ~.
\end{align}
Thus, for any input transmitter $\rho(s,r)$ we can always choose a pure
transmitter $\psi(s,r)=|\psi\rangle\langle\psi|$ such that $\chi\lbrack
\Phi|\rho(s,r)]\leqslant\chi\lbrack\Phi|\psi(s,r)]$. As a result, the
maximization in Eq.~(\ref{readCAP}) can be restricted to pure transmitters
$\psi(s,r)$.

\subsection{Triviality of the unconstrained version of the
capacity\label{APPtrivial}}

Here we provide a simple sketched proof showing that unconstrained quantum
reading capacity simply equals the whole data stored in the marginal cell of
the memory.

Let us consider a pure transmitter in the tensor-product form $\psi
(s)=\psi^{\otimes s}$ at the input of a memory with marginal cell $\Phi
=\{\phi_{x},p_{x}\}$. At the output of the cell the arbitrary state is given
by%
\begin{equation}
\rho_{x}(s)=[\phi_{x}(\psi)]^{\otimes s}~,
\end{equation}
where $\phi_{x}(\psi)$\ is the single-copy output state. Since the quantum
channels $\phi_{x}$ are different, for any pair $\phi_{x}$ and $\phi
_{x^{\prime}}$ there is at least an input (pure) state $\psi$ such that
\begin{equation}
F[\phi_{x}(\psi),\phi_{x^{\prime}}(\psi)]=\varepsilon<1~. \label{F_xxp}%
\end{equation}
For the sake of simplicity, let us assume that this state $\psi$ is the same
for all the channels, i.e., the Eq.~(\ref{F_xxp}) holds for any $x\neq
x^{\prime}$. Then, by exploiting the multiplicativity of the fidelity under
tensor product states, we get%
\begin{equation}
F[\rho_{x}(s),\rho_{x^{\prime}}(s)]=\varepsilon^{s}~,
\end{equation}
for any $x\neq x^{\prime}$. Now, since this quantity goes to zero for
$s\rightarrow+\infty$ we have that the multi-copy output states $\rho_{x}(s)$
become asymptotically orthogonal. This implies that%
\begin{equation}
\chi\lbrack\Phi|\psi(s)]=\chi(\{\rho_{x}(s),p_{x}\})\rightarrow H(X)~.
\end{equation}
The proof can be easily extended to the weakest case where Eq.~(\ref{F_xxp})
holds for different input states $\psi_{i}$ where $i=0,\cdots,k-1$ for a
$k$-ary variable $X$. In this general case, for high values of $s\gg k$, we
consider input states%
\[
\psi(s)=\psi_{0}^{\otimes s_{0}}\otimes\cdots\otimes\psi_{k-1}^{\otimes
s_{k-1}}%
\]
where $s_{0}+\cdots+s_{k-1}=s$. It is easy to check that the output states
become asymptotically orthogonal.


\begin{thebibliography}{99}
\bibitem {Assisted}Here we formulate the assisted readout for the specific
case of the optical memories (bosonic channels). It is understood that the
assisted readout can also be formulated for other kinds of memories
(non-bosonic) under different physical constraints.
\bibitem {NoteEnergy}Here we fix the signal energy effectively irradiated over
the cell since this is the weakest constraint we can impose. Clearly this
means that we do not impose any constraint on the reference modes. An
alternative, but more demanding energy constraint, corresponds to fixing the
mean number of photons of the global input state, i.e., including both signal
and reference modes.
\bibitem {Transmitters}Note that the notation adopted here for the
transmitters is a bit different from the one used in
Refs.~\cite{Pirs,PirsSUPP}. However, as in Refs.~\cite{Pirs,PirsSUPP}, the
notation is a compact way to indicate input states with variable Hilbert spaces.
\bibitem {capp}Technically speaking the quantity in Eq.~(\ref{readCAP}) is a
\textquotedblleft single-cell\textquotedblright\ capacity\ since there is no
entanglement distributed among the different cells. However, the probing of
each cell includes the use of a generally entangled reference system which is
shared by Alice and Bob. For this reason, it represents an \textquotedblleft
entanglement-assisted\textquotedblright\ capacity. The capacity becomes
\textquotedblleft unassisted\textquotedblright\ by setting $r=0$\ in
Eq.~(\ref{readCAP}).
\end{thebibliography}
\end{document}